\documentclass[prl,twocolumn,preprintnumbers,amsmath,amssymb,secnumarabic,aps,superscriptaddress]{revtex4}
\usepackage{dcolumn}
\usepackage{bm}
\usepackage[normalem]{ulem}
\usepackage{color}
\usepackage{float}
\usepackage{subfig}

\usepackage{amstext}
\usepackage{amssymb}
\usepackage{epsfig,amssymb}

\begin{document}


\title{Internal microstructure driven turbulence enhancement of fluids }

\author{G. Sofiadis}
\affiliation{Department of Mechanical
Engineering, University of Thessaly, 38334, Volos, Greece.}
\author{I. E. Sarris}
\email{sarris@uniwa.gr}
\affiliation{Department of Mechanical
Engineering, University of West Attica, Athens, Greece.}

\date{\today}

\begin{abstract}
Fluids with internal microstructure like dense suspensions, biological and polymer added fluids, are commonly found in the turbulent regime in many applications. Their flow is extremely difficult to be studied as microstructure complexity and Reynolds number increase, even nowadays. This bottleneck is novelty and efficiently treated here by the micropolar theory. Our findings support that when denser microstructure occurs turbulence is intensified by the chaotic rotation of internal fluid's elements. Unlike in Newtonian fluids, shear stress is now decreased, and viscous and microstructure interrotational stresses increase near the walls and mark the turbulence intensification.

\end{abstract}

\maketitle

In the usual Newtonian fluid flow, shear is the dominant mechanism for turbulence production, however, as  the fluid departs from the Newtonian regime, its microstructure is found to alter the usual turbulent characteristics  i.e. to result in drag increase \cite{Picano:2015} or decrease~\cite{Dallas:2010}. Turbulence intensification or reduction is not reported to be only due to viscosity increase is such cases, but rather than due to the  presence of internal elements like its microstructure, added polymer strings or microparticles. Phenomena like the increase of  effective viscosity, turbophoresis,  shear thickening or thinning are associated with the existence of fluid's internal microstructure \cite{Wagner:2009,Stickel:2005,Reeks:1983,Picano:2013}.    In the case of suspensions, it is reported that as the volume fraction of additives increases, their induced stress is increased and results in shear stress reduction. 

The usual turbulence production mechanism is affected or even dominated that way by  the internal elements induced turbulence. As particle volume fraction increases  their turbulent interaction increases and sustains high total drag. This is found not to be related to higher friction velocity, but to the increased viscosity of dense suspensions \cite{Picano:2015}.
More specific, turbulence mechanism in case of dense suspensions does not only vary in terms of production but location as well. It has recently been proved~\cite{Costa:2016} that when particles dominate the flow (become larger than the smaller turbulent scales), near wall turbulence production becomes significant. This near-wall turbulence mechanism has been attributed to an effective suspension viscosity, but is still not fully understood.  A similar near-wall mechanism has been found to arise in general, when fluid microstructure is taken into consideration. This peculiar turbulence enhancement mechanism seems to be directly connected to a microstructure stress tensor.

Polymer and viscoelastic fluids  may also contain deformable micro-structures. The major advantage of such fluid turbulent flows is the reduced drag due to polymeric additions. Toms~\cite{Toms:1949} firstly observed this pressure drop decrease when polymers are added in water or other solvents. Later in an experimental study of polymer dilute pipe flows, Giles and Pettit~\cite{Giles:1967} agreed that in similar types of flows, like Toms~\cite{Toms:1949},  polymer additives  in low concentrations decrease friction. Experiments in tubes were also conducted by White and McEligot~\cite{White:1970} for different types of polymers in deionized water solutions. Their results showed a polymer-concentration dependent Reynolds number for transition to turbulence.  Moreover, in a recent paper, Rosti and Brandt~\cite{Rosti:2020} measured with the same methodology that polymer additives in dense suspensions  may lead to drag decrease.

Dense suspensions or   other fluids with internal microstructure  need special numerical attention of their motion and interaction of  rigid internal additives, whereas  in significant high Reynolds numbers it is extremely difficult to be used for simulations. This bottleneck is treated here by the micropolar theory which is able to mathematically describe a wide range of fluids spanning from dense suspensions, liquid crystals,  blood and other polymeric-like fluids consisting of internal microstructure~\cite{Eringen:1965}. The class of equations that describes the micropolar theory is attractive due to its capability to include an asymmetric and couple stress tensor in a rather simple way~\cite{Eringen:1965,Mitarai:2002}. The symmetric stress tensor arises in macroscopic continuum theory when certain degrees of freedom are assigned to the fluid, such as internal spin, leading to a generalization of the incompressible, $\nabla\cdot u =0$, \ Navier-Stokes equation~\cite{Eringen:1965,Mitarai:2002,Luk:1999,Condiff:1964}, as: \vspace{-5pt}
\begin{equation}\label{masseq}
\frac{\partial u}{\partial t}+ \left(u \cdot \nabla \right) u = - \nabla P + \frac{1}{Re}\nabla^{2} u + \frac{m}{Re} \nabla\times \omega 
\end{equation} \vspace{-20pt}
\begin{equation}\label{eneq}
\frac{JN}{m} \left[ \frac{\partial \omega}{\partial t} + (u \cdot \nabla)\omega\right] = \frac{1}{Re} \nabla^{2} \omega + \frac{N}{Re} \nabla\times u - \frac{2 N}{Re} \omega \vspace{-5pt}
\end{equation}
where, $u$ and $\omega$ are the linear and angular velocity vectors, respectively, and $t$ and $P$ stand for time and pressure, respectively. 

In this Letter, Eqs.~(\ref{masseq})-(\ref{eneq}) are solved  for the usual  turbulent channel flow test-bed cases of~\cite{Tsukahara:2005,Kim:1986,Moser:1999}. We adopt the set of constitutive equations that have been originally proposed by Eringen~\cite{Eringen:1965}, as an extension of the classical Navier-Stokes equation for fluids with microstructure. Direct numerical simulations (DNS) of the turbulent channel flow at various $Re$ and $m$ values are performed to demonstrate the key role of the micropolar effect  at turbulence generation.
Turbulent channel flow is a well documented test-bed for this type of flows and large databases of turbulent statistics are available for Newtonian and non-Newtonian fluids \cite{Dallas:2010,Picano:2015,Tsukahara:2005,Kim:1986,Moser:1999}. The immediate advantage of the present model is that a mesh of about 50 times smaller is needed to perform the dense suspensions simulations at the same Reynolds number as in ref.~\cite{Picano:2015}. Furthermore, both Eqs.~(\ref{masseq})-(\ref{eneq}) can be uniformly solved in a Eulerian frame, with no need to treat a discrete phase.

The linear and angular velocities, time and pressure at the above equations are made dimensionless by selecting the characteristic quantities $U_{0}$, $\frac{U_{0}}{\delta}$, $\frac{\delta}{U_{0}}$, and $\rho U_{0}^2$, respectively, where  $U_{0}$  is the constant mean velocity,  $2\delta$ is the height of the channel and $\rho$ is the density of the fluid which is considered to be homogenous. At Eqs.~(\ref{masseq})-(\ref{eneq}) four non-dimensional quantities are recovered, i.e. the modified bulk Reynolds number, $Re=\rho\frac{U_{0}2\delta}{\mu + \kappa}$, with $\mu$ being the molecular and $\kappa$ the micropolar fluid viscosities, the so-called vortex viscosity parameter, $m=\frac{\kappa}{\mu + \kappa}$, the dimensionless microrotation parameter, $J=\frac{j}{\delta^2}$, where $j$ is the microinertia of the fluid, and the so-called spin gradient viscosity parameter, $N=\frac{\kappa \delta^2}{\gamma}$, where $\gamma$ is the material coefficient of the fluid~\cite{Eringen:1966,Cheikh:2019}.  
Throughout this study, the  spin gradient viscosity and the microrotation parameters are kept constant at $N=8.3\times 10^{4}$ and $J= 10^{-5}$ that are usual magnitudes for biological flows~\cite{Ariman:1974}, while  $m$ is varied between 0 and 0.9. This is chosen because of the straightforward connection between the linear and angular momentum equations through $m$.
Fluids near $m=0$ behave similar to the Newtonian one, while when  $m=0.9$ a microstructure dominated fluid is considered.
Increase of $m$ in the present simulations, under constant bulk Reynolds numbers, imply the simultaneously reduction of  molecular viscosity so that the total viscosity of the fluid to remain the same. 

\begin{figure}
  \begin{center}
    \includegraphics[width = 200pt, height=100pt] {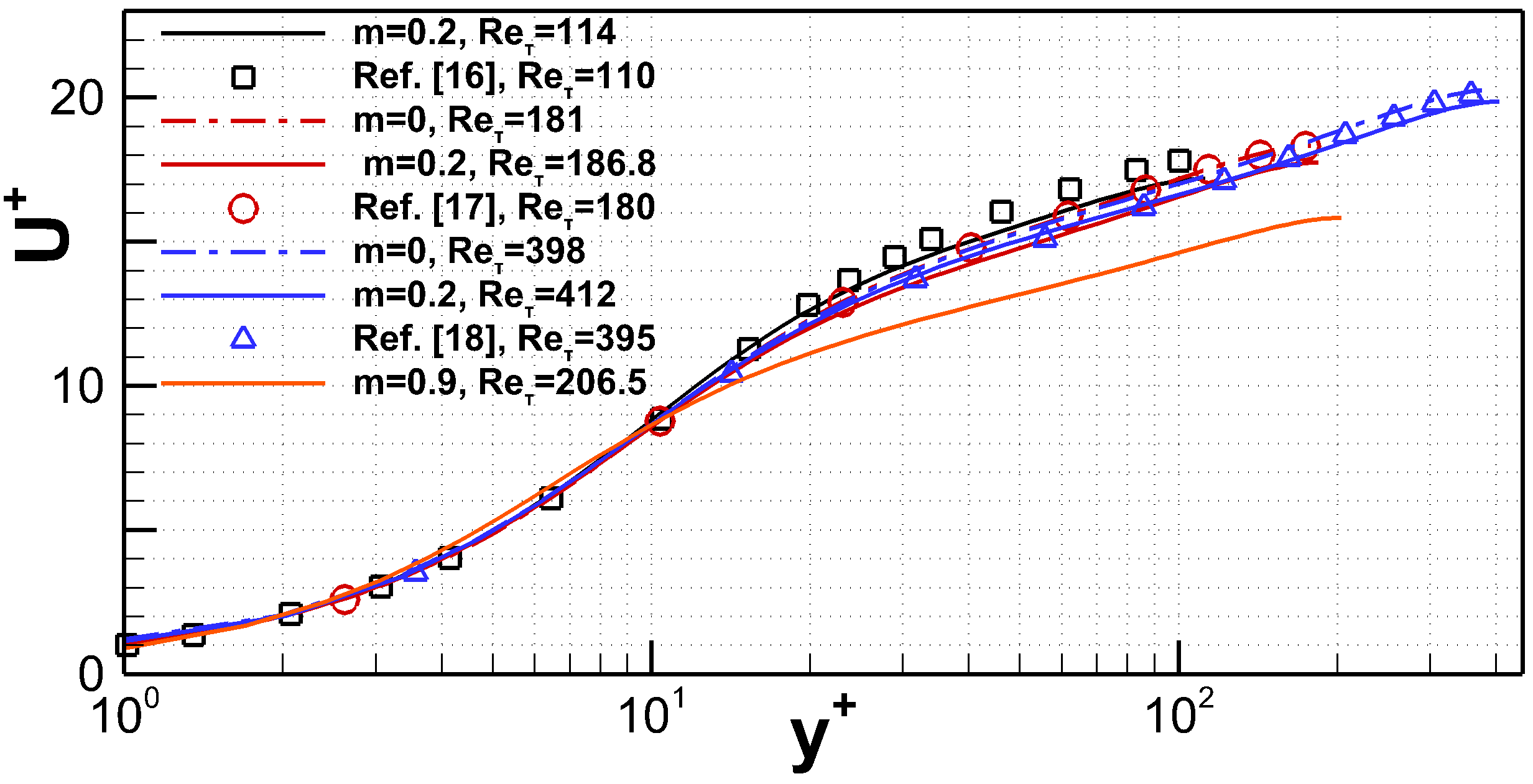}
    \caption{Mean velocity distributions  along $y^{+}$ for various Newtonian cases and the case of $m=0.9$.}
    \label{fig:Uplus}
\vspace{-20pt}
  \end{center}
\end{figure}

Initially, successful experiments are made in the Newtonian regime to compare our numerical facility against others~\cite{Moser:1999}. As can be seen in  Figure~\ref{fig:Uplus} for the temporal and wall-normal plane averaged velocity $U^{+}$ along the wall-normal direction $y^{+}$, both normalized by the shear velocity $u_{\tau}=\sqrt{\tau_{w}/\rho}$,  where $\tau_{w}$ is the shear stress and $^{+}$ indicates inner (wall) units. Cases in the Newtonian range of $Re=3300-13800$ are considered here, or equivalent in the range $Re_{\tau}=110-395$,  where the Newtonian friction Reynolds number is based on the shear velocity defined by   $Re_{\tau}= \frac{\rho u_{\tau} \delta}{\mu + \kappa}$. In all these Newtonian cases, the channel size and grid arrangement is kept equal as in the DNS study of Moser et al.~\cite{Moser:1999}. Moreover, the accuracy of the micropolar part of the model is already verified in recent works~\cite{Karvelas:2020a,Karvelas:2020b} against other numerical and experimental results.  
\begin{figure}
  \begin{center}
    \includegraphics[width = 200pt, height=100pt] {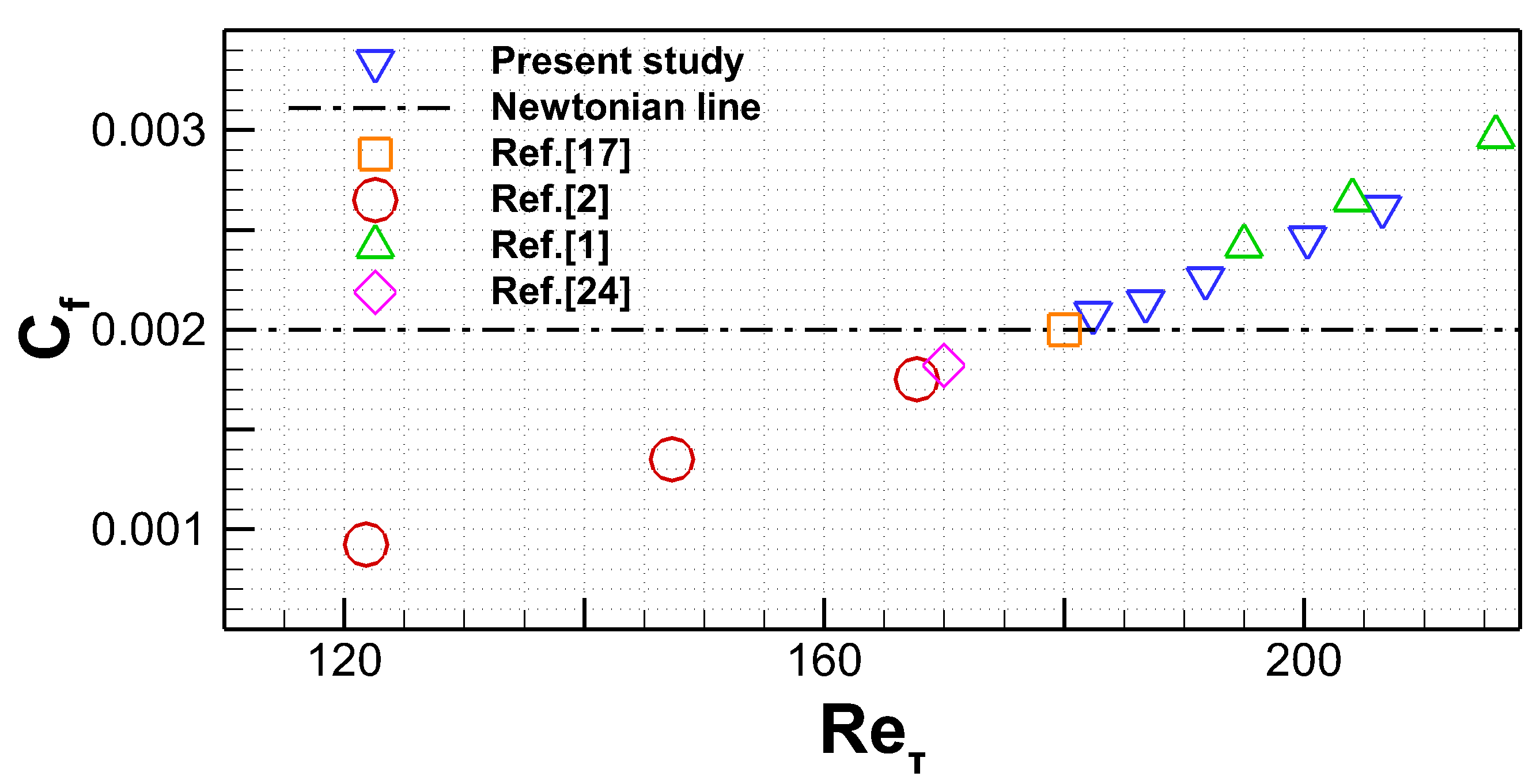}\vspace{-5pt}
    \caption{Skin friction coefficient from different DNS results of micropolar, particulate and polymer additive flows at  $Re_{\tau}=180$ based on the Newtonian case. }\vspace{-10pt}
    \label{fig:drag}
\vspace{-15pt}
  \end{center}
\end{figure}

As can be seen in Fig.~\ref{fig:Uplus} for example at $Re_{\tau}=180$, the increase of the vortex viscosity parameter from zero in the Newtonian case to $m=0.9$ cause slope decrease of   $U^{+}$  and consequently, pressure drop increase and turbulence intensification due to the strongly interrotating microstructure. The $m$ increase is also connected to the increase of  $Re_{\tau}$ by  8\% at most in the $Re_{\tau}=
395$ case due to increased $u_{\tau}$ as the microstructure effect is more significant near the walls, as  will be discussed later.  The rate of turbulence enhancement as connected by the presence of  microstructure can be efficiently presented  in the friction factor, $C_{f}=\tau_{w}/\frac{1}{2} \rho U^2_0$, schema of Fig.~\ref{fig:drag}. There, the friction factor variation of various non-Newtonian cases is compared against the present micropolar results as $m$ increases. In the case of flows containing polymer additives~\cite{Dallas:2010} or small heavy solid particles~\cite{Dritselis:2008} friction is found to decrease in lower values than in the Newtonian one, however, in the case of dense suspensions~\cite{Picano:2015} as well as in the present case of internal microstructure, friction factor
is increased, and thus, turbulence is enhanced with volume fraction and vortex viscosity increase, respectively.

\begin{figure}
  \begin{center}
    \includegraphics[width = 120pt, height=40pt] {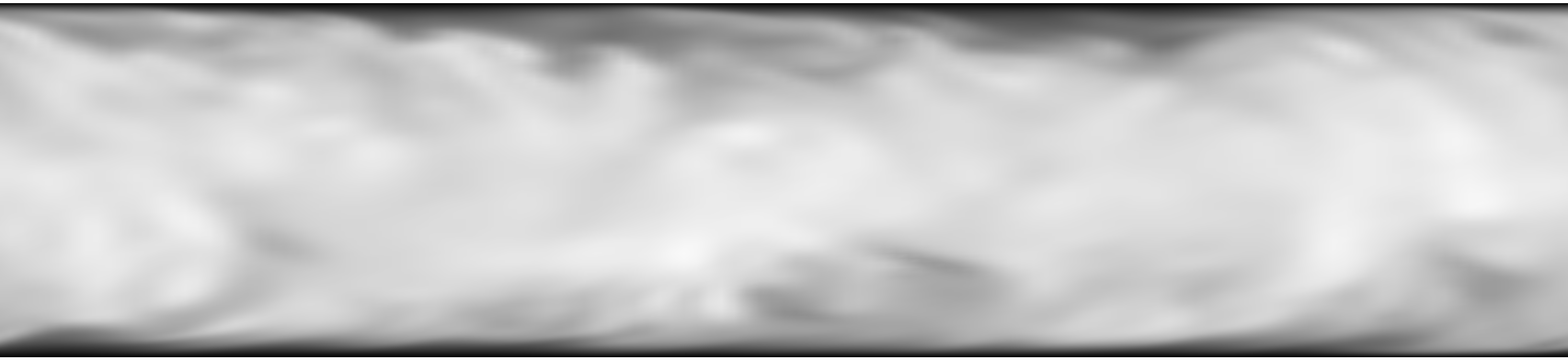}
    \includegraphics[width = 120pt, height=40pt] {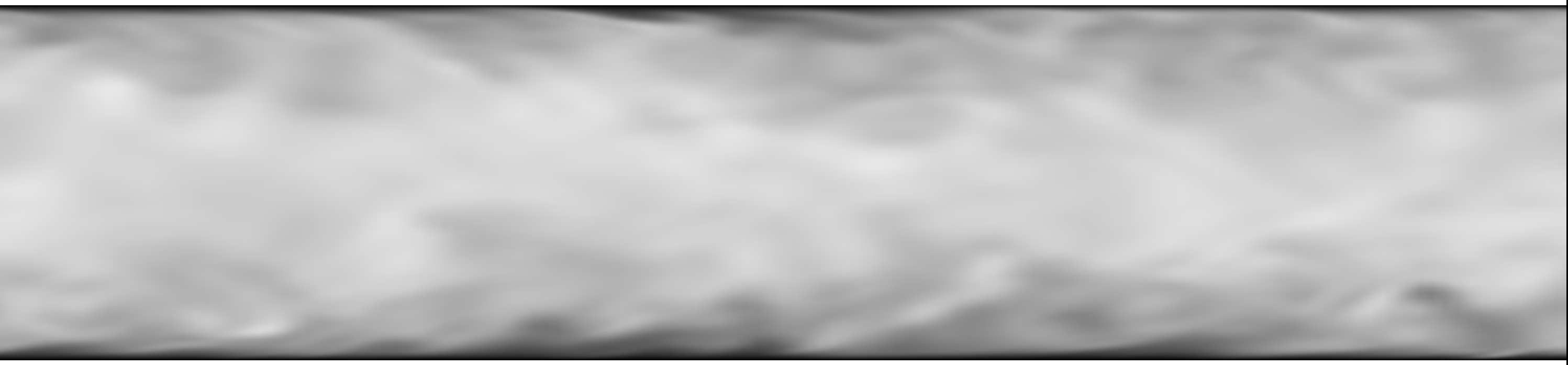}
    
     \includegraphics[width = 120pt, height=80pt] {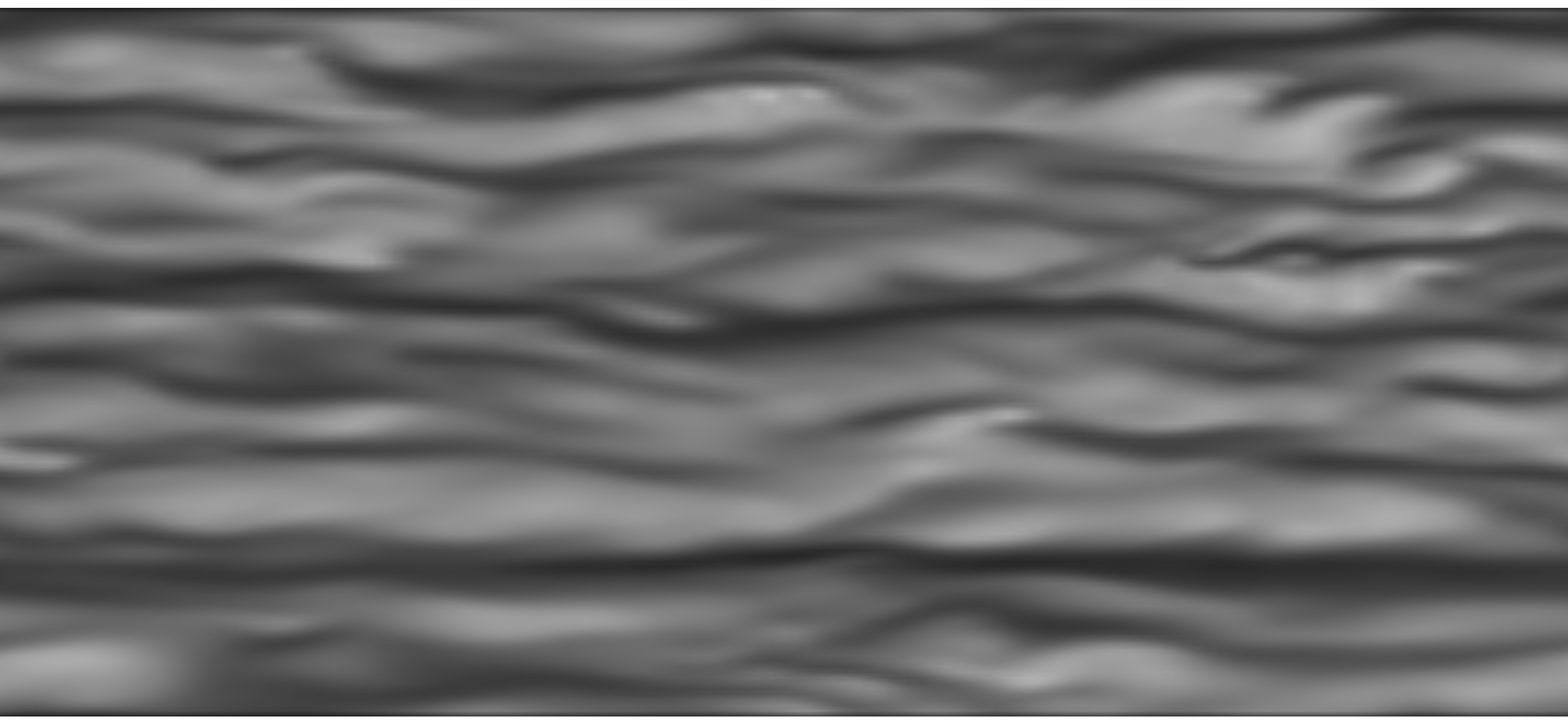}
     \includegraphics[width = 120pt, height=80pt] {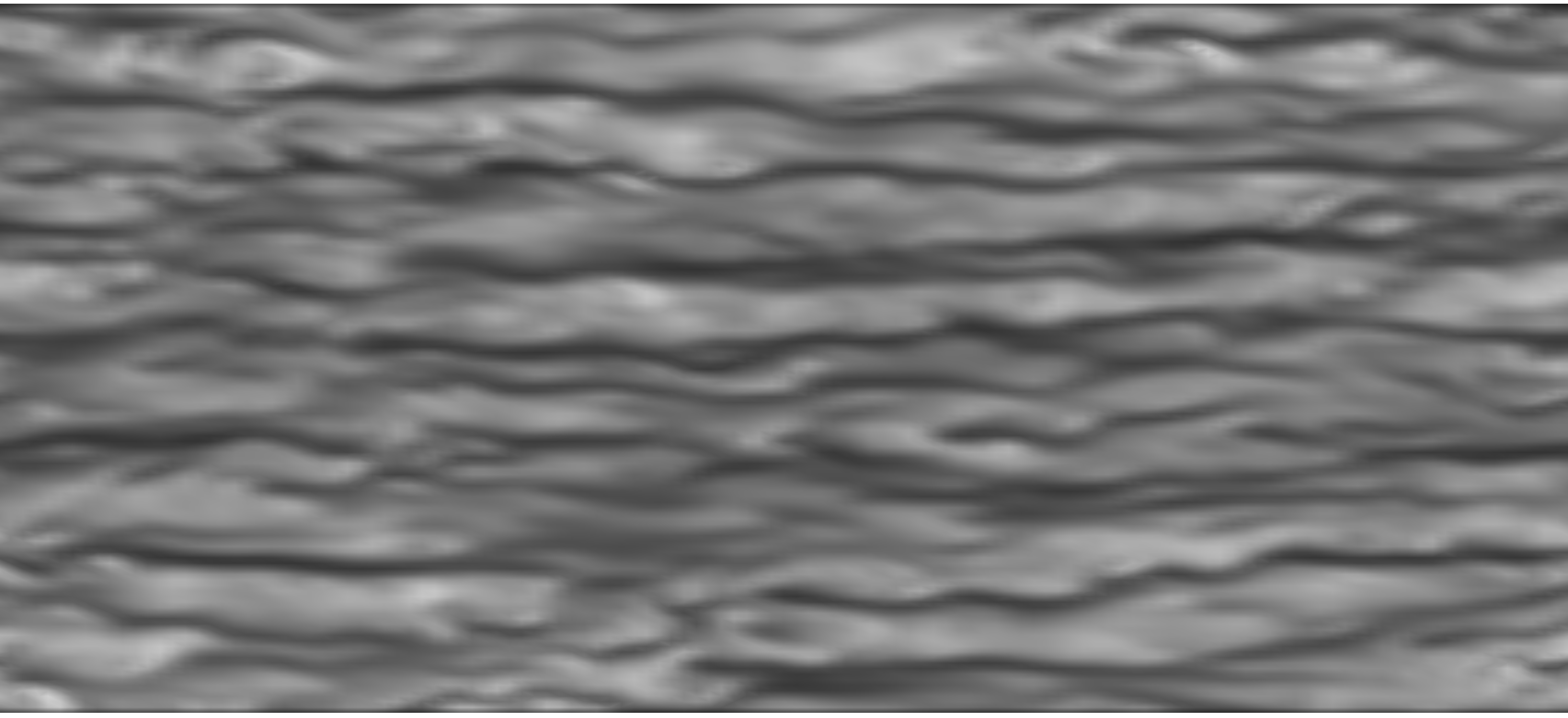}
    \caption{Streamwise velocity contour snapshots at a streamwise plane (up) and at a wall-normal plane (bottom) at $y^+=1$  for a Newtonian fluid at  $Re_{\tau}=180$  (left) and as $m=0.9$  (right).}\vspace{-10pt}
    \label{fig:con}
\vspace{-20pt}
  \end{center}
\end{figure}

Turbulence enhancement due to microstructure is not only found to exist quantitatively in the above figures, but more importantly qualitatively in the different forms between the flow structures. This is made visual in the streamwise velocity contour snapshots aligned with the $x-y$ and $x-z$ planes of Fig.~\ref{fig:con}. It is observed that finer structures are encountered for $m = 0.9$ than for $m = 0$ in the $x-y$ plane that is a strong indication of shear thickening and the tendency for more  isotropic turbulence. Furthermore, the stronger turbulent mixing is evident in the $x-z$ plane, for example at $y^{+}=1$, where finer and isotropic structures are recorded for $m= 0.9$.

\begin{figure}[t]
  \begin{center}
    \includegraphics[width = 120pt, height=100pt] {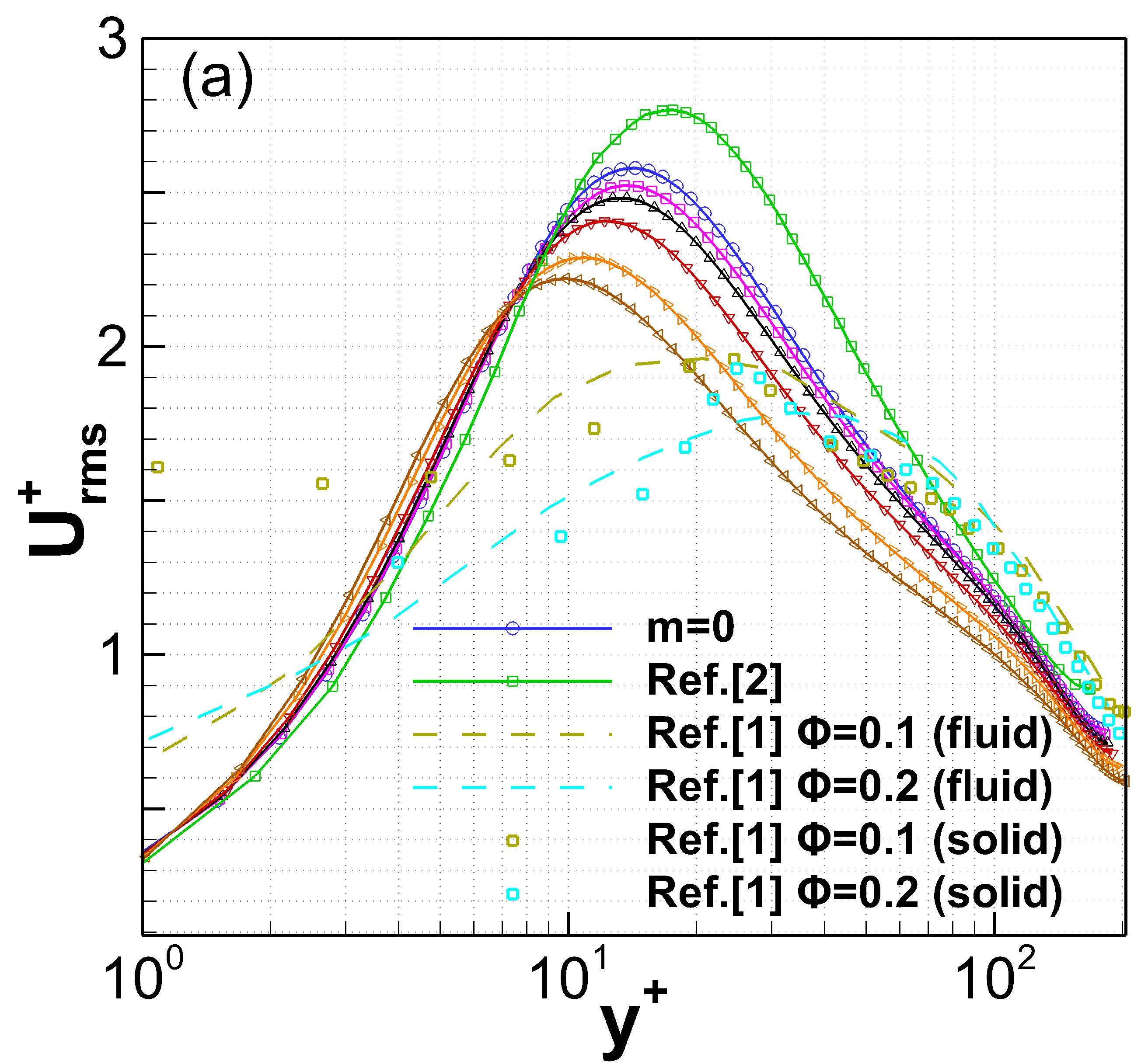}
    \includegraphics[width = 120pt, height=100pt] {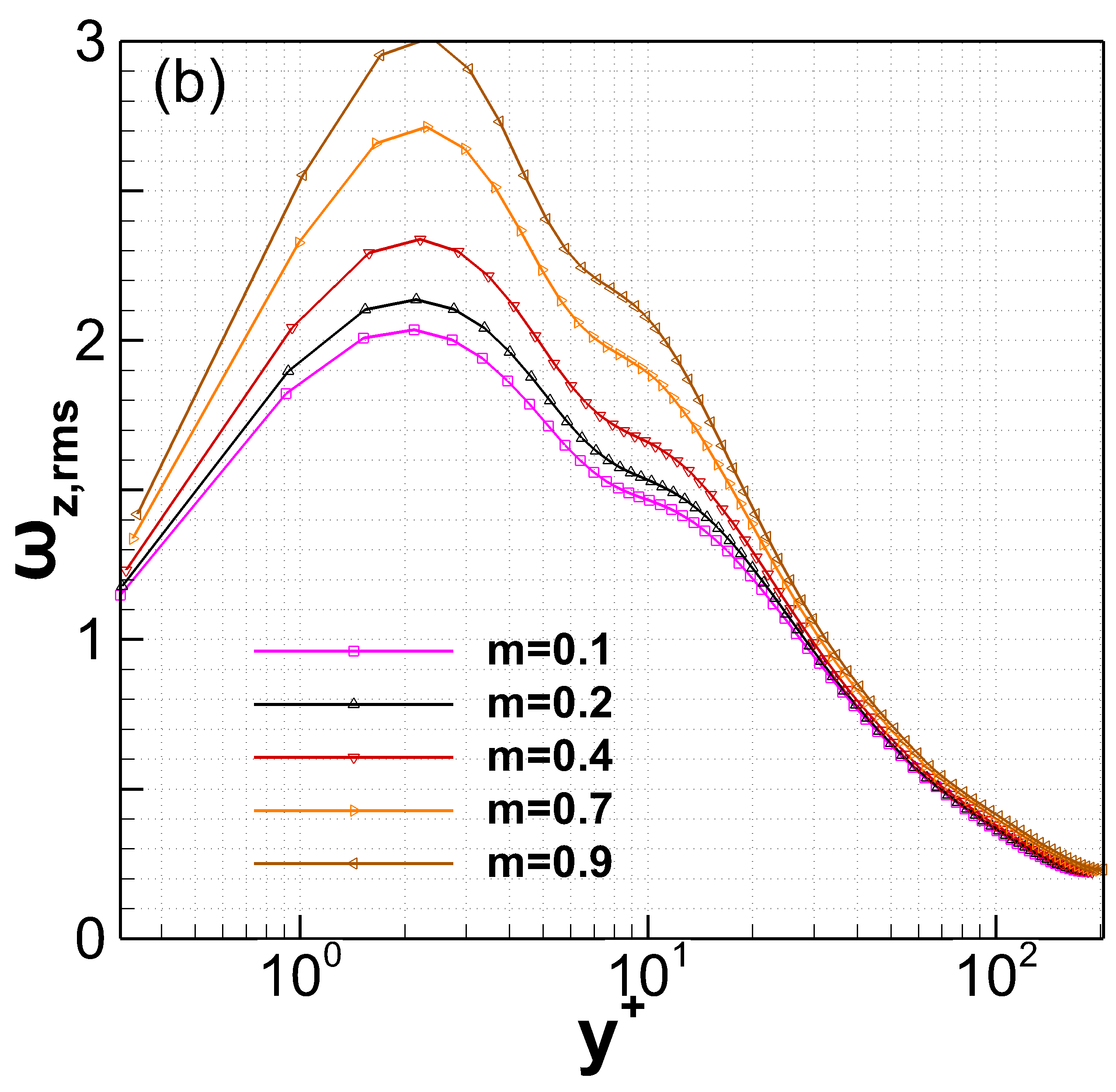}
    \caption{Root-mean-square  velocity fluctuations normalized by the wall-shear velocity: a) the linear streamwise component $U^+_{rms}$, and b) the angular spanwise component $\omega^+_{z,rms}$.}\vspace{-5pt}
    \label{fig:rms}
\vspace{-20pt}
  \end{center}
\end{figure}

Turbulent statistics, like the velocity root-mean-squared fluctuations  that are shown in Fig.~\ref{fig:rms}, also support the turbulence intensification finding due to microstructure. There, some of the present results  are compared against polymeric and dense suspension cases at the same Newtonian  flow of $Re_{\tau}=180$. As $m$ increases, the peak value of the streamwise fluctuation component of the velocity, $U^+_{rms}$, is reduced and approaches the wall. Thus, the peak from $y^+ \approx 20$ at $m= 0$ goes at about $y^+ \approx 10$ for   $m= 0.9$. Simultaneously, the reduction of peak value  indicates the turbulent homogenization trend of internal microstructure due to increasingly turbulent fluctuations of the angular velocity, $\omega^+_{z,rms}$, as $m$ increases, see Fig.~\ref{fig:rms}b. The micropolar fluctuation  $\omega^+_{z,rms}$ that mostly interacts with  $U^+_{rms}$ is found to  approach its peak value at about $y^+ \approx 10$ for   $m= 0.9$. As compared to the polymer additive case~\cite{Dallas:2010}, an opposite trend is
found due to its  shear thinning  nature.

Moreover, the present results are in alignment with the dense suspension case of ref.~\cite{Picano:2015} in two major points, although the visual differences. Firstly, the peak values of  $U^+_{rms}$ are reduced due to shear thickening and this reduction is proportional to particles volume fraction increase, and secondly, despite that the peak value of   $U^+_{rms}$ of the fluid phase at ref.~\cite{Picano:2015} departs from the walls with volume fraction increase, its particulate counterpart approaches the walls like in the present results. The peak value of the particulate root-mean-squared fluctuation velocity of ref.~\cite{Picano:2015} is found to be of the same order of magnitude as its fluid part, thus, turbulence intensification is fed by the particulate induced turbulence in dense suspensions.       

Similarly here, the internal microstructure is responsible for the turbulence enhancement as can be seen in the stress balance of the micropolar fluids flow equations. In turbulent wall-bounded fluid flows with internal microstructure,  the total stress is balanced by Reynolds shear, viscous and microstructure stresses as: \vspace{-5pt}
\begin{equation}\tau_{t}=\tau_{s}+\tau_{v}+\tau_{m}\vspace{-5pt}\end{equation}
where, $\tau_{s}=-\overline{u^{+'} v^{+'}}$ and $\tau_{v}=\overline{\frac{d U^{+}}{d y^{+}}}$ represent the shear and viscous stresses, respectively, while the micropolar stress due to microstructure is $\tau_{m} = m\overline{\omega^+_z}$, $u'$ and $v'$ are the streamwise and wall-normal velocity fluctuations, respectively, and bar indicates  temporal and wall-normal plane average. 

\begin{figure} 
  \begin{center}
    \includegraphics[width = 120pt, height=100pt] {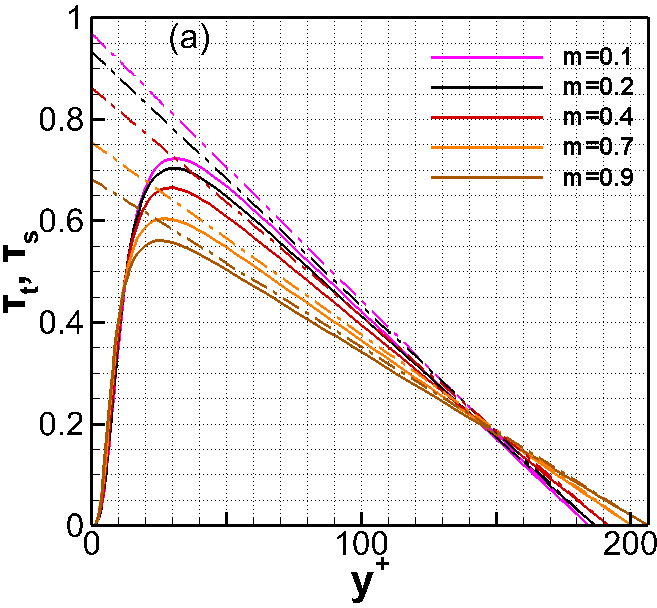}
    \includegraphics[width = 120pt, height=100pt] {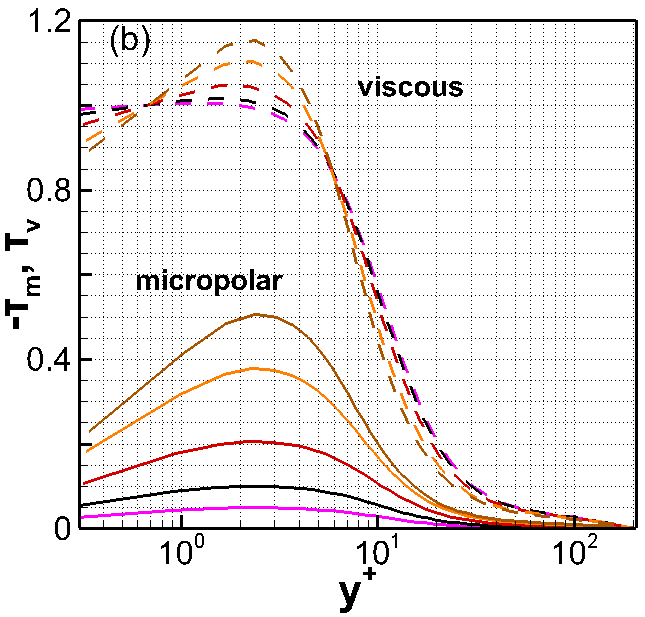}
    \includegraphics[width = 120pt, height=100pt] {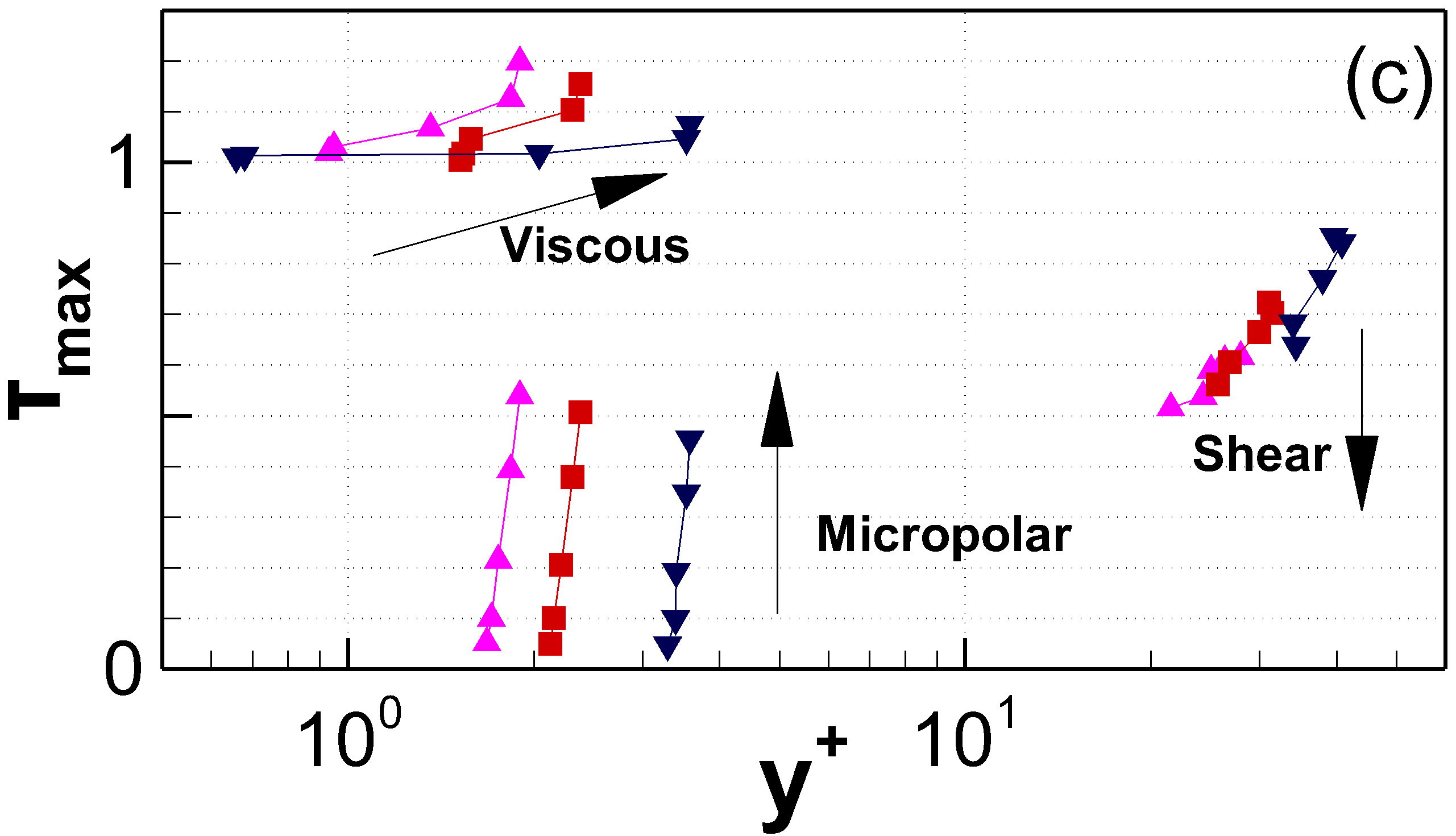}
    \vspace{-5pt}\caption{Stress  profiles for various $m$ along  $y^{+}$ at  $Re_{\tau}=180$  for $m = 0$: a) total and shear stresses,  b) viscous and micropolar stresses, and c) locations of the peak values of all stresses. Arrows indicate the increase of $m$, Lines with $\blacktriangle$ at $Re_b=3300$, with $\blacksquare$ at $Re_b=5600$, and with $\blacktriangledown$ at $Re_b = 13800$.}
    \label{fig:tsb}
\vspace{-20pt}
  \end{center}
\end{figure}

Reynolds stress is the usual mechanism of turbulence production, but here is peculiarly reduced with $m$ increase  as can be observed in Fig.~\ref{fig:tsb}. Similar  shear stress decrease is encountered in dense suspensions, polymer flows and flows under the effect of external forces, like external magnetic fields~\cite{Lee:2001}. Shear reduction is replaced by the simultaneous increase of both, the viscous and micropolar stresses near the walls at $y^+ \lesssim\ 40$.
This new turbulence mechanism ignited by the internal microstructure is reported for the first time here. Thus now, the micropolar stress alters  the dominance of viscous stress near the walls and, as $m$ increases, the magnitude of  $\tau_{m}$ is getting comparable to that of the  viscous one. The micropolar stress  is found to be of the same nature as the viscous one, i.e. rises only very close to the wall, while far from it drops to zero and mostly affected by the total viscosity. The peak value of the micropolar stress is found to be located inside the viscous sublayer at  $y^+ \approx 2-3$  for the lower Reynolds number of  $Re_b=3300$ almost independently of $m$, and with an tendency to departs from the wall as Reynolds number increases.
This decrease of shear stress
 is responsible also for the reduction of  total stress, similarly to~\cite{Lee:2001}, due to the homogenization effect of the internal microstructure. 

The near walls new induced turbulence mechanism through microstructure  is of similar nature as the usual vorticity induced turbulence mechanism of the Newtonian fluids (now act together) that excite streaks  structures near the walls~\cite{Schoppa:2002}.
In fact, this is evident from Fig.~\ref{fig:tsb} where the profiles of the mean angular velocity, i.e. $m \overline{\omega^+_z}$,  is of similar shape to the fluid spanwise friction vorticity profile, i.e.     
$\overline \Omega_{z}=\overline{\frac{d  U^{+}}{d y^{+}}}$. It is found easily from the present statistics, and also proved mathematically, that $2\overline{\omega_z} =\overline \Omega_{z}=\overline{\frac{d  U^{+}}{d y^{+}}}$. The difference however of the micropolar stress production is that it is an active term capable to sustain and enhance turbulence, and not a passive one as the usual vorticity is.

In summary, the micropolar set of equations has been applied to investigate the internal microstructure effect in a turbulent fluid flow. Computational experiments have been conducted at relative low Re channel flow and results for different vortex viscosity ratios have been compared against other non-Newtonian  fluids flows. Little is known so far for this type of  turbulent flows, which ranks this study among pioneer ones. Results exhibit shear thickening, drag increase and enhanced turbulence as the vortex viscosity increases. By further analysis, it has been shown that the main mechanism for turbulence generation in this case is the micropolar stress, as it increases close to the walls when  vortex viscosity, and thus microstructure density, increases. The concept of micropolar turbulence formation could lead to a better understanding of turbulence mechanics in more general type of fluids.

The authors are grateful for the support of the Greek Research and Technology Network (GRNET) for the computational time granted in the National HPC facility ARIS.

\end{document}